\documentclass[prd,aps,twocolumn,superscriptaddress,amsmath,amssymb]{revtex4-2}
\usepackage{epsfig}
\usepackage{graphicx}
\usepackage[dvipsnames]{xcolor}
\usepackage{upgreek}
\usepackage{hyperref}
\hypersetup{
	colorlinks=true,
	pdfborder={0 0 0},
	citecolor=blue,
	linkcolor=blue,
	urlcolor=blue
}
\usepackage[normalem]{ulem}

\begin{document}

\title{Role of non-gaussian quantum fluctuations in neutrino entanglement}

\author{Denis Lacroix } \email{lacroix@ijclab.in2p3.fr}
\affiliation{Universit\'e Paris-Saclay, CNRS/IN2P3, IJCLab, 91405 Orsay, France}

\author{A.\ B.\ Balantekin}
\affiliation{%
Department of Physics, University of Wisconsin--Madison,
Madison, Wisconsin 53706, USA
}%

\author{Michael J.\ Cervia}
\affiliation{Department of Physics, The George Washington University,
Washington, District of Columbia 20052, USA
}
\affiliation{Department of Physics, University of Maryland, 
College Park, Maryland 20742, USA
}

\author{Amol V.\ Patwardhan}
\affiliation{%
SLAC National Accelerator Laboratory,
Menlo Park, CA 94025, USA $\>\>\>\>\>\>$
}%

\author{Pooja Siwach}
\affiliation{%
Department of Physics, University of Wisconsin--Madison,
Madison, Wisconsin 53706, USA
}%

\date{\today}
\begin{abstract}
The flavor evolution of neutrinos in environments with large neutrino number densities is an open problem at the nexus of astrophysics and neutrino flavor physics. Among the many unanswered questions pertaining to this problem, it remains to be determined whether neutrino-neutrino coherent scattering can give rise to nontrivial quantum entanglement among neutrinos, and whether this can affect the flavor evolution in a meaningful way. To gain further insight into this question, here we study a simple system of two interacting neutrino beams and obtain the exact phase-space explored by this system using the Husimi quasi-probability distribution.
We observe that the entanglement induced by the coupling leads to strong delocalization in phase-space with largely non-Gaussian quantum fluctuations. The link between the neutrino entanglement and quantum fluctuations is illustrated using the one- and two-neutrino entropy. In addition, we propose an approximate phase-space method to describe the interacting neutrinos problem, where the exact evolution is replaced by a set of independent mean-field evolutions with a statistical sampling of the initial conditions. The phase-space approach provides a simple and accurate method to describe the gross features of the neutrino entanglement problem. Applications are shown using time-independent and time-dependent Hamiltonians in the non-adiabatic regime.         
\end{abstract}


\maketitle

\section{Introduction}

In compact object astrophysical environments such as core-collapse supernovae and binary neutron star mergers, neutrinos play a potentially significant role in the dynamics and nucleosynthesis (e.g.,~\cite{Ful92, Dua11, Xio20}), making it vital to understand their flavor evolution. These environments are characterized by extremely high number densities of neutrinos, which can lead to a multitude of collective flavor oscillation phenomena driven by neutrino-neutrino coherent scattering (e.g.,~\cite{Dua09,Dua10,Cha16,Tam21} and references therein). 
One intriguing facet of this problem is the possibility of neutrinos experiencing quantum entanglement due
to neutrino-neutrino coherent scattering \cite{Bel03, Fri03a, Fri03b, Fri06, Cer19, Rra19, Rog21, Pat21, Bal22}. 
The presence of entanglement could modify neutrino oscillation patterns, inducing departures from the \lq\lq mean field\rq\rq\ approximation, wherein such entanglement is explicitly forbidden. 
However, just like many-body systems in other areas of physics, describing a system of mutually interacting neutrinos 
is known to become intractable rapidly as the particle number increases.

Much recent effort has focused on providing an accurate description of neutrino many-body systems, especially on exploring  their novel quantum behavior.
The problem is particularly difficult to treat,
first because of its many-body nature and second because the Hamiltonian 
should \emph{a priori} be considered time-dependent beyond the adiabatic limit. 
Recently this problem was addressed using 
Bethe ansatz techniques \cite{Peh11,Bir18,Pat19} and the tensor network approach \cite{Rog21,Cer22}. Using  
quantum computers is also being explored \cite{Hal21,Yet22,Kum22,Ill22,Ami22}. These many-body techniques apply feasibly for small numbers of neutrinos or neutrino beams or for time-independent or slowly evolving Hamiltonians. 
  
In the present study, we take a different starting point. The evolution of interacting neutrinos and 
their entanglement is analyzed in phase-space. Phase-space methods can be very useful in discussing quantum information (e.g.,~\cite{Flo22}). Many criteria to discuss quantum entanglement are based on the second moments of some observables \cite{Sim20,Dua00, Man03,Gio04,Guh04,Hyl06}.  
However, these criteria cannot resolve non-Gaussian entanglement between bipartite systems \cite{Wee12,Ser17}. Here, we construct directly the Husimi 
function associated with the interacting particles and analyze its connection to quantum entanglement. We further show 
that the exact phase-space evolution can be qualitatively mimicked
by starting from a statistical sampling of initial conditions and evolving them independently
as prescribed in Refs.~\cite{Ayi08,Lac14}.     

In the next section, we recall some ingredients of the two-neutrino beams problem and its exact solution as well as its approximate mean-field solution. 
In Sec.~\ref{sec:entanglement}, we show the explicit connection between the quantum fluctuations and the quantum entanglement of neutrinos. Section \ref{sec:husimi} discusses the Husimi quasi-probability distribution of
the neutrinos belonging to one of the beams, which gives
insight into these quantum fluctuations. Finally, in Sec.~\ref{sec:smf}, we present an approach based on a set of 
mean-field evolutions with initial random fluctuations able to describe approximately the complexity of the quantum fluctuations and entanglement 
for the two-neutrino beams problem.


\section{The two neutrino beams setup}

\subsection{The two-neutrino beams Hamiltonian}
\label{sec:2beam}

The phase-space analysis is made here in the so-called ``two-beam geometry" (e.g., \cite{Fri03b,Fri06,Mar21,Xio21}). We assume two flavors and consider an ensemble of $i=1,\ldots,N$ neutrinos where each neutrino is described by a two-level system associated with the creation operators $(a^\dagger_{1,i}, a^\dagger_{2,i})$. The corresponding single-particle states $|1, i \rangle$ and 
$|2,i\rangle$ are the neutrino's mass-basis eigenstates. We assign to each neutrino the quasi-spin operators $(j^{i}_{\pm}, j_{z}^{i})$ defined as
\begin{eqnarray}
j^i_+ &=& a^\dagger_{1, i} a_{2,i}, ~ j^i_z =  \frac{1}{2} \left(  a^\dagger_{1, i} a_{1,i} - a^\dagger_{2, i} a_{2, i} \right). \label{eq:quasispin1}
\end{eqnarray}
From these components, the spin vectors $\vec j_i = (j^i_x, j^i_y, j^i_z)$ are defined with $j^i_\pm = j^i_x \pm i j^i_y$. 
These operators together with the number operator $n_i = a^\dagger_{1, i} a_{1,i} +a^\dagger_{2, i} a_{2, i} $ obey standard SU(2) algebra.  
In the two-beam approximation, the neutrinos are split into two subsystems  called beams $A$ and $B$ with $N_A$ and $N_B$ particles, respectively,  
that interact through 
the Hamiltonian \cite{Bal07,Mar21}: 
\begin{eqnarray}
\frac{H}{\mu} &=& \frac{\Omega}{2} \vec B \cdot \left( \vec J_A - \vec J_B \right) + \frac{2}{N} \vec J_A \cdot \vec J_B , \label{eq:2beam} 
\end{eqnarray}  
where $\vec J_{A/B} = \sum_{i \in A/B} \vec j_i $ denotes the total quasi-spin operators of $A$ and $B$. Such an approximation has been 
widely used as a test-bench for more general (e.g., \cite{Dua06,Dua10,Bal06,Bal07}) neutrino oscillation problems, both in the mean-field approximation, and in many-body treatments.  In Eq.~(\ref{eq:2beam}), $\vec B$ 
equals $(0,0, -1)$ in the mass basis. 
Following Ref.~\cite{Mar21}, we assume that all neutrinos in a given beam have identical momenta (magnitude and direction) and initial flavor.
The initial state of the system is then given by a Slater determinant denoted by $| \Psi \rangle = | \Omega_A \rangle \otimes | \Omega_B \rangle$. Here, we have:
\begin{eqnarray}
|\Omega_A \rangle = | \theta_A, \phi_A \rangle = \prod_{i=1}^{N_A} a^\dagger_{A,i} | 0_A\rangle \label{eq:coherent}
\end{eqnarray} with 
\begin{eqnarray}
a^\dagger_{A,i} &=& \cos \left( \frac{\theta_{A}}{2}\right)a^\dagger_{1,i} +  \sin \left( \frac{\theta_{A}}{2}\right) e^{i \phi_{A}} a^\dagger_{2,i}, \label{eq:coherent2}
\end{eqnarray}
where the $(\theta_A, \phi_A)$ angles defined the transformation from the mass to flavor single-particle basis.
Such initial states correspond to a SU(2) coherent state for the present problem \cite{Bal07,Gil78,Zha90,Gaz95,Vie95}. 
The state $|\Omega_B \rangle$ is defined similarly using the angles $(\theta_B, \phi_B)$ to depict the neutrino composition of system $B$. 

\subsection{Exact evolution}

To obtain the exact solution, we use the method of Ref.~\cite{Xio21} that was further developed in Ref.~\cite{Mar21}
and extended to the three beams geometry in Ref. \cite{Rog22}. Using the symmetry of the initial state together with the conservation 
laws induced by the simplified Hamiltonian, we can decompose the exact solution at all times as 
\begin{eqnarray}
| \Psi (t)\rangle = \sum_{m_A, m_B} C_{m_A, m_B}(t) | m_A  , m_B \rangle. \label{eq:psit}
\end{eqnarray}
Here, we use the compact notations  $| m_A , m_B \rangle  = | J_A, m_A \rangle \otimes |J_B, m_B \rangle $
with $ | J_{A/B}, m_{A/B} \rangle$ denoting the standard angular momentum eigenstates. The numbers of neutrinos 
$N_A$ and $N_B$ in the subsystems $A$ or $B$ are constants of motion. These numbers are linked to the total spin by the relationships  
$J_{A/B} = N_{A/B}/2$ together with $m_{A/B}=-J_{A/B}, \cdots, J_{A/B}$.
Note that, the exact solution is obtainable here, because we start from an initial state that is fully symmetric with respect to the exchange of neutrinos within either subsystem. Because of this symmetry, only states with highest multiplet value for each subsystem appear in Eq.~\eqref{eq:psit}. This simplification reduces considerably the number of components to consider
and renders the exact problem numerically tractable. 
If the symmetry with respect to the permutation of indices 
is broken initially in one of the subsystems, then all multiplets should be considered, and consequently the method would become much more difficult---if not impossible---to solve numerically as the number of particles increases.  

To obtain the exact solution of the problem, it is useful to realize that the Hamiltonian is block diagonal in the subspace where 
$M= m_A + m_B = {\rm constant}$  \cite{Xio21}.  This fact could indeed be proven by first rewriting the Hamiltonian as:
\begin{eqnarray}
\frac{H}{\mu} &=& - \frac{\Omega}{2} \left( J^A_z - J^B_z \right) + \frac{2}{N} J^A_z J^B_z  \nonumber \\
&&+ \frac{1}{N} \left( J^A_+ J^B_- + J^A_- J^B_+  \right) . \label{eq:ham2}
\end{eqnarray}   
The first two terms are diagonal in the basis and we have:
\begin{eqnarray}
\langle m_A, m_B | - \frac{\Omega}{2} \left( J^A_z - J^B_z \right)| m_A, m_B \rangle&=& - \frac{\Omega}{2}(m_A - m_B), 
\nonumber \\
\langle m_A, m_B |   \frac{2}{N} J^A_z J^B_z    | m_A, m_B \rangle&=&  \frac{2}{N} m_A m_B .  \nonumber 
\end{eqnarray}
The last term in Eq. (\ref{eq:ham2}) gives the transitions 
\begin{eqnarray}
J^A_+ J^B_- | m_A, m_B \rangle &=& U_{m_A, m_B} | m_A+1, m_B-1 \rangle  \label{eq:pm} \\
J^A_- J^B_+ | m_A, m_B \rangle &=&  D_{m_A, m_B} | m_A-1, m_B+1 \rangle  \label{eq:mp} 
\end{eqnarray}
with 
\begin{eqnarray}
U_{m_A, m_B} &=& \sqrt{\left[ J_A(J_A+1) - m_A(m_A+1)  \right] } \nonumber \\
&\times& \sqrt{\left[ J_B(J_B+1) - m_B(m_B-1)  \right]} , \nonumber \\
D_{m_A, m_B} &=& \sqrt{\left[ J_A(J_A+1) - m_A(m_A -1)  \right] }\nonumber \\
&\times& \sqrt{\left[ J_B(J_B+1) - m_B(m_B + 1)  \right]}. \nonumber
\end{eqnarray}
From these expressions, it becomes evident that the Hamiltonian only couples states of the same $M$ and 
is block-diagonal in this representation. Here we consider a sufficiently small number of neutrinos to numerically diagonalize the Hamiltonian 
in each block with constant $M$:
\begin{eqnarray}
M &=& - \frac{N_A+N_B}{2}, \cdots,  \frac{N_A+N_B}{2}. \nonumber
\end{eqnarray} 

In practice, once all eigenvalues and eigenvectors are obtained separately for the blocks corresponding to different values of $M$, 
the exact solution of the problem can be computed provided that we have the initial values of the coefficients 
$C_{m_A, m_B}$. For the specific initial condition considered 
in this work, these coefficients are given by:
\begin{eqnarray}
C_{m_A, m_B}(0) &=& \sqrt{C^{N_A/2 + m_A}_{N_A} } [{\rm c}_A]^{\frac{N_A}{2}+ m_A} [ {\rm s}_A e^{i \phi_A}]^{\frac{N_A}{2}- m_A} \nonumber \\
&\times& \sqrt{C^{N_B/2 + m_B}_{N_B} } [ {\rm c}_B]^{\frac{N_B}{2}+ m_B} [s_B e^{i \phi_B }] ^{\frac{N_B}{2} - m_B},  \nonumber 
\end{eqnarray}
with the notation ${\rm c}_{A/B} = \cos \left(\theta_{A/B} /2 \right)$ and ${\rm s}_{A/B} = \sin \left(\theta_{A/B} /2 \right)$. The present method does not have 
any specific difficulty provided that the maximal value of $M$, i.e. number of neutrinos $N_{A/B}$ is not too large. A more specific discussion 
on practical aspects and Hilbert space size of the present method can be found in Ref. \cite{Mar21}.  

The exact total density operator corresponds to a pure state density $D(t)= | \Psi(t) \rangle \langle \Psi(t) |$ at all 
times, where $| \Psi(t) \rangle$ is given by Eq.~\eqref{eq:psit}. Starting from this total density, one can access the reduced 
density of each neutrino beam using $D_{A/B} (t) ={\rm Tr}_{B/A} D(t)$. 
Focusing on the system $A$ and using Eq.~(\ref{eq:psit}), we obtain:
\begin{equation}
D_A(t)=\sum_{m_A, m'_A} | m_A \rangle\left[ \sum_{m_B} C_{m_A, m_B} (t) C^*_{m'_A, m_B} (t)   \right] \langle m'_A |,  \label{eq:dat}
\end{equation} 
from which any observable related to the subsystem $A$ can be numerically estimated.   

\begin{table}
  \centering
    \begin{tabular}{| l |c| c |c|c|c |}
    \hline
Mode      & $\Omega$ & $\theta_A$  & $\phi_A$ & $\theta_B$ & $\phi_B$ \\
     \hline \hline 
Bipolar & $0.5$ & $\pi-0.2$ & $\pi$ & $0.2$ & $0$ \\
\hline 
Precession  &  $1.2$ &  $0.5978067$ & $0$ & $0.2175694$ & $0$ \\
    \hline 
    \end{tabular}
  \caption{Two sets of parameters that are used in the illustration (Bipolar and Precession modes \cite{Mar21}).}
  \label{tab:paramcte}
\end{table}

\begin{figure}[htbp] 
\includegraphics[scale=0.35]{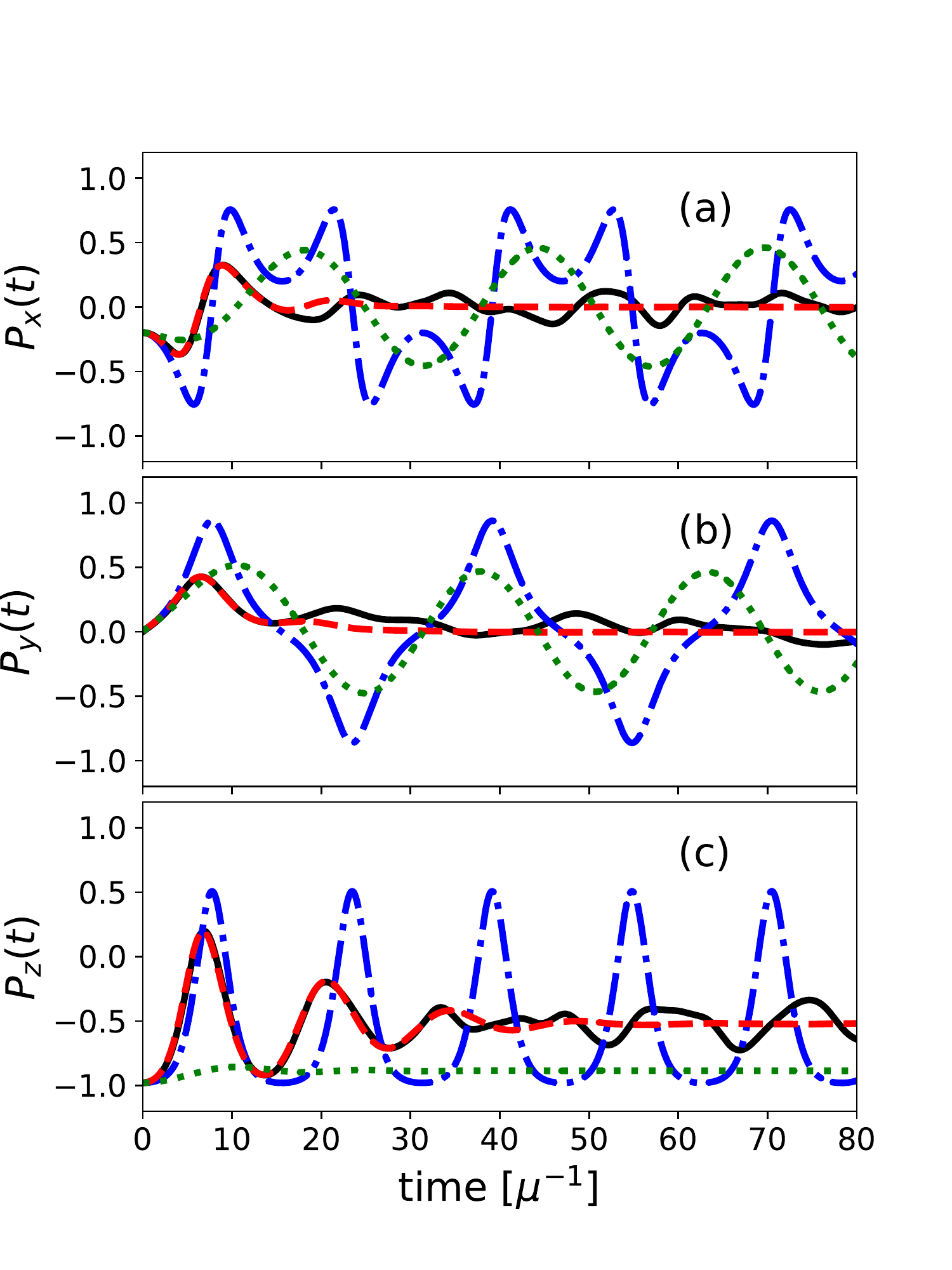} \\
\includegraphics[scale=0.35]{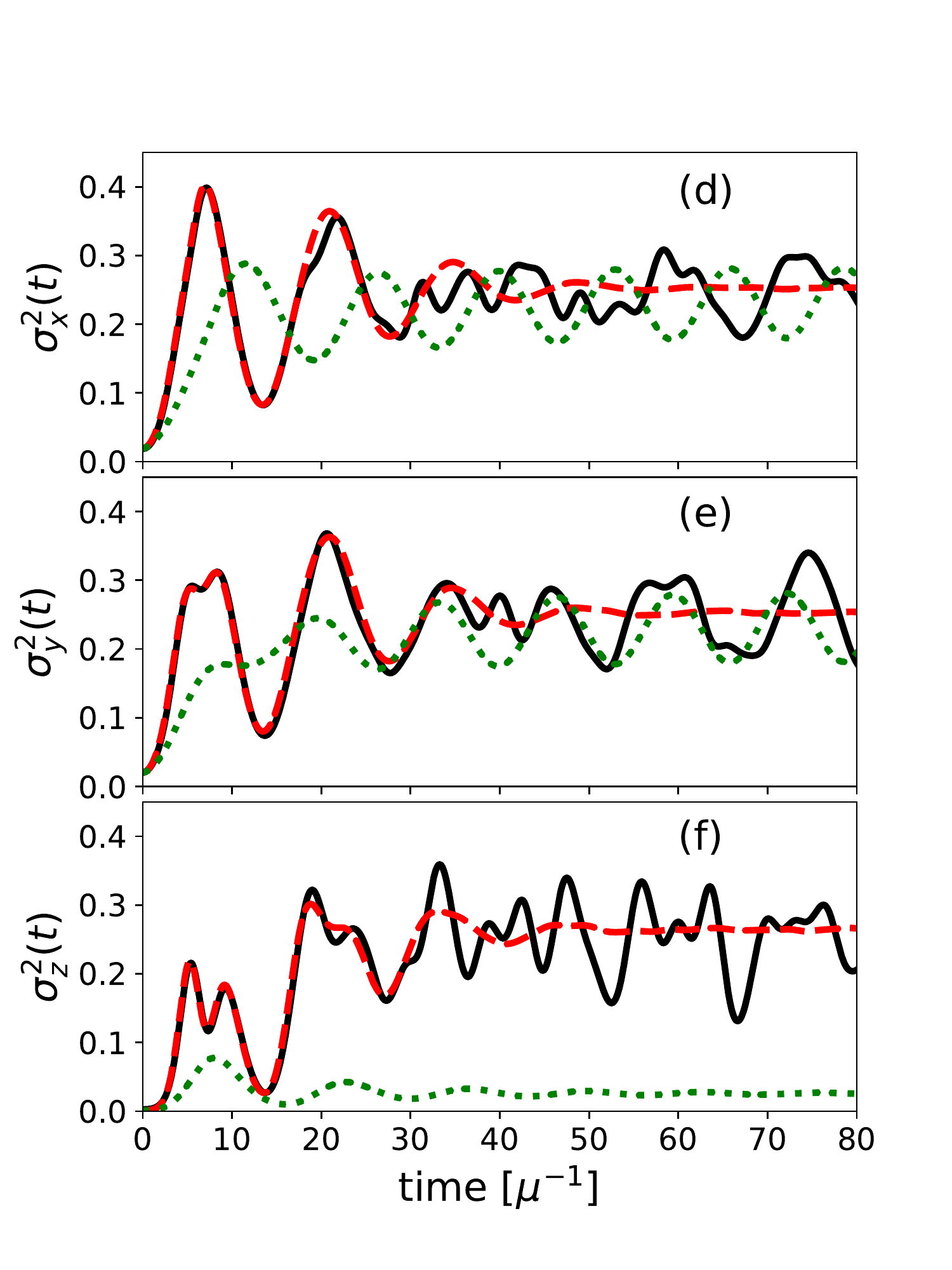} 
    \caption{Illustration of the evolution of the polarization components $ \vec P_A$ (top) and their quantum fluctuations (bottom) for the Bipolar Symmetric case  with $N_A=N_B =50$ as a function of time for the exact (black solid), mean-field (blue dot-dashed), and phase-space method (red dashed). 
The Bipolar Symmetric case corresponds to an initial state described by a Slater determinant (see Sec.~\ref{sec:2beam}) with the different angles 
recalled in Table \ref{tab:paramcte}.     
In the bottom panel, the mean-field fluctuations are constant in time and remain equal to their initial values (not shown).
The green dotted curve corresponds to the time-dependent Hamiltonian case discussed in Sec.~\ref{sec:timedep}.}
    \label{fig:BipSymP123}
\end{figure}

Examples of exact evolutions of the polarization components $ \vec P_A  = 2 \langle \vec J_A \rangle/N_A$ are given in Fig. \ref{fig:BipSymP123} for the so-called \lq\lq Bipolar symmetric\rq\rq\ case of Ref. \cite{Mar21} with initial parameters listed in Table \ref{tab:paramcte}. 
We also present in panel (b) the associated second moments defined as 
\begin{eqnarray}
\sigma^2_{A,\alpha} = \frac{4}{N_A^2} \left[ \langle J^2_{A,\alpha}   \rangle - \langle J_{A,\alpha} \rangle^2 \right]
\end{eqnarray}
with $\alpha=x,y,z$. 
In the same vein as previous studies (e.g.,~\cite{Pat19, Rra19, Cer19, Pat21, Cer22, Peh11, Bir18, Xio21, Mar21, Rog21, Rog22, Bal22}), any signatures of quantum entanglement between the neutrinos and the consequent departures from the mean-field behavior are expected to be imprinted in the evolution of these first and second moments $P_{A,\alpha}$ and $\sigma^2_{A,\alpha}$.

In the following discussion, we will mainly focus our analysis to the bipolar symmetric case. However, we also studied other cases discussed in Ref.~\cite{Mar21}. The conclusions we draw below apply in all cases we studied. For the sake of completeness, we also show additional illustrations of exact evolutions for alternative sets of parameters and initial conditions in Appendix \ref{app:example}.

\subsection{Mean-Field approximation}

Exact solutions of coupled neutrino beams can be obtained in very few cases, i.e., when the numbers of beams and of neutrinos in each beam are both small
and when the Hamiltonian is time-independent, which is not the case for neutrinos emitted from supernovae. For these reasons, as already mentioned in the
introduction, extensive efforts are being made to develop many-body approximations grasping the physics of neutrino oscillation as much as possible. 
The simplest approximation is certainly the mean-field theory. This approximation has a limited predictive power because it is unable to properly account for the 
two-body correlations. An advantage, which is rarely underlined, is that it easily accommodates a time-dependent Hamiltonian even in the non-adiabatic regime. As we will see below, the mean field 
will play the role of the Gaussian phase-space distribution that will serve as a reference. It also will be used to design an approximate phase-space method able to accurately describe beyond mean-field effects. We briefly recall here the mean-field equations of motion (EOMs) for the 
model case considered in the article.                

The mean-field approximation for the ``two-beam geometry'' has been derived in several works (see, for instance, Refs.~\cite{Mar21,Xio21}).
We only give here the main steps leading to the equations of motion of the polarization components that are solved numerically. In particular, we derive the mean-field EOMs from the Ehrenfest theorem applied to the quasi-spin components:
\begin{eqnarray}
i \hbar \frac{\mathrm{d} \langle \vec J_{A,B} \rangle }{\mathrm{d}t} &=& \langle [ \vec J_{A/B} , H ] \rangle.
\end{eqnarray}  

A straightforward manipulation of the quasi-spin operators leads to the set of exact coupled equations: 
\begin{eqnarray}
\left\{ 
\begin{array}{c}
\displaystyle \frac{\mathrm{d}}{\mathrm{d}t} \langle \vec J_A\rangle =  + \frac{\Omega}{2} \vec B \wedge \langle \vec J_A\rangle +   \frac{2}{N} \langle \vec J_B \wedge 
\vec J_A \rangle,  \\
\\
\displaystyle \frac{\mathrm{d}}{\mathrm{d}t} \langle \vec J_B \rangle = -  \frac{\Omega}{2} \vec B \wedge \langle \vec J_B\rangle +   \frac{2}{N} \langle \vec J_A \wedge 
\vec J_B \rangle .
\end{array}\right.  \label{eq:jajb}
\end{eqnarray}  
Solving these equations requires us to write and solve the equations of motion of the moments $\langle J^{A}_\alpha J^{B}_{\beta} \rangle$ with $\alpha, \beta=x,y,z$  where the coupling to higher moments of the quasi-spin appears. This leads to the equivalent of the so-called BBGKY hierarchy 
\cite{Bog46,Bor46,Kir46,Bon16}. 
 
The mean-field theory assumes $\langle J^{A}_\alpha J^{B}_{\beta}\rangle \simeq \langle J^{A}_\alpha \rangle \langle  J^{B}_{\beta}  \rangle$,
which is a strong approximation for the quantum fluctuations. 
Introducing the expectation values of the polarization components of both systems, we end up with the six coupled EOMs:
\begin{eqnarray}
\left\{ 
\begin{array}{l}
\displaystyle  \dot P^A _x =  + \frac{\Omega}{2}  P^A _y + x_B C^{AB}_x   \\  
\\
\displaystyle  \dot P^A _y =  - \frac{\Omega}{2}  P^A _x +  x_B C^{AB}_y   \\  
\\
\displaystyle  \dot P^A _z =  +  x_B C^{AB}_z    \\
\\
\displaystyle  \dot P^B _x =  - \frac{\Omega}{2}  P^B _y - x_A C^{AB}_x   \\  
\\
\displaystyle  \dot P^B _y =  + \frac{\Omega}{2}  P^B _x -  x_A C^{AB}_y   \\  
\\
\displaystyle  \dot P^B _z =  -  x_A C^{AB}_z    
\end{array}
\right. 
\label{eq:mf}
\end{eqnarray}
with 
\begin{eqnarray}
C^{AB}  &=&  \langle \vec P_B \rangle \wedge \langle \vec P_A \rangle, ~{\rm and}~
x_{A/B} = \frac{N_{A/B} }{N} .\nonumber
\end{eqnarray}

For the specific initial state considered in the main text, the above EOMs are solved with the initial conditions:
\begin{eqnarray}
P^A_x &=& \sin(\theta_A) \cos(\phi_A)  ,  ~~P^B_x =  \sin(\theta_B) \cos(\phi_B),\nonumber \\
P^A_y &=& \sin(\theta_A) \sin(\phi_A) , ~~P^B_y = \sin(\theta_B) \sin(\phi_B) , \nonumber \\
P^A_z &=& \cos(\theta_A), ~~\hspace*{1.2cm} P^B_z = \cos(\theta_B) . \nonumber 
\end{eqnarray}

As shown in Fig.~\ref{fig:BipSymP123} and further illustrated in Appendix \ref{app:example}, the mean-field 
theory is able to describe the short time evolution of one-body observables but fails to reproduce the exact evolution 
at longer time even for one-body quantities. Several general (connected) considerations can be made about the origin 
of the failure of mean-field: (i) This approximation generally poorly treats quantum fluctuations and their effects on one-body observables during the time evolution; (ii) A related aspect is that the quantum entanglement induced by the two subsystems coupling is essentially neglected. This absence actually becomes evident by noting that the mean-field approximation is equivalent to assuming that the wave function remains separable during the evolution, i.e. 
$| \Psi (t) \rangle = | \Omega_A(t) \rangle \otimes | \Omega_B(t) \rangle$ where $| \Omega_{A/B}(t) \rangle$ are Slater determinants. (iii) Mean-field theory leads in general to significant underestimation of quantum fluctuations. 
Besides, it implicitly assumes Gaussian fluctuations in phase-space which turns out to be a too drastic approximation for the neutrino oscillations problem (see below). 

In the following section, we clarify below the connection between quantum fluctuations and quantum entanglement between neutrinos (i.e., items (i) and (ii)), demonstrating 
that the proper description of quantum fluctuations is a prerequisite to describe entanglement in neutrino physics. We then 
make a complete characterization of quantum fluctuations by performing a phase-space analysis of the two neutrinos beam exact evolution. 
Such analysis is not only useful to understand the departure from a mean-field picture, but also a strong guidance for 
proposing an efficient many-body approach for the neutrino oscillation problem.   

\section{Connection between quantum fluctuations and two-particle entanglement}
\label{sec:entanglement}

To trace the connection of quantum fluctuations in phase-space to entanglement, we compute the von-Neumann entropy for either a given neutrino $i$ or a pair of neutrinos $(i \neq j)$ from subsystem $A$ directly in the reduced Fock space. For this purpose, we use the technique developed in Ref.~\cite{Rob21}. 

We first construct explicitly the one-body reduced density and associated entropy of a single neutrino $i$ belonging to the subsystem $A$. We suppose that the reduced density $D_A(t)$ is known (given for instance by Eq.~\eqref{eq:dat} for the exact case), 
and we would like to construct the reduced one-neutrino density
$R^{(i)}_1$. In order to do so, we consider the full Fock space basis for subsystem $A$. A state in this basis can generically be written as:
\begin{eqnarray}
| n_{1,2}, n_{1,1} , \cdots ,  n_{i,2}, n_{i,1} , \cdots , n_{N_A, 2},  n_{N_A, 1} \rangle , \nonumber
\end{eqnarray}        
where $n_{i,1/2} = 0,1$ depending on whether the corresponding single-particle state $|{i,1}\rangle$ or $|{i,2}\rangle$ is occupied.  
$i$ refers here to the given neutrino.
We therefore see that the subspace associated with a given neutrino contains four states denoted hereafter
by $\{ | 00\rangle_{i},  | 01\rangle_{i},   | 10\rangle_{i} ,  | 11\rangle_{i} \}$, using the shorthand notation 
$ | 00\rangle_{i} = | n_{i,2} = 0 , n_{i,1} = 0 \rangle $. 
The one-neutrino density associated with the neutrino $i$ is then obtained by tracing the total density over the other neutrinos, i.e. 
\begin{eqnarray}
R^{(i)}_1 &=& {\rm Tr}_{1, \cdots , i-1, i+1, \cdots, N_A} D_A(t).
\end{eqnarray}   
To obtain an explicit form of the reduced density, we use the following properties:
\begin{eqnarray}
{}_i \langle n_2 n_1 | R^{(i)}_1  | n_2 n_1 \rangle_i  &=& {\rm Tr} \left( |n_2 n_1 \rangle_i \langle n_2 n_1|_i D_A \right)  .
\end{eqnarray} 
We then re-express the operator in the trace using the expressions of the spins associated with neutrino $i$ given in Eq.~\eqref{eq:quasispin1}. 
These identities give the correspondence: 
\begin{eqnarray}
|01\rangle_i \langle 01| &=& \frac{1}{2}(1_i +2 j^i_z), \nonumber \\
|10\rangle_i \langle 10| &=& \frac{1}{2}(1_i -2 j^i_z), \nonumber \\
|10\rangle_i \langle 01| &=&  \hat j^i_+ , ~~~
|01\rangle_i \langle 10| =  \hat j^i_- . \nonumber 
\end{eqnarray}

An important property of our system is that there is strictly one neutrino $i$ which prevents any contributions from the states $|00\rangle_i$ 
and $|11\rangle_i$. Using these properties, we finally deduce that the reduced density is given by: 
\begin{eqnarray}
R^{(i)}_1 & = & \frac{1}{2}
 \left[ 
\begin{array}{cccc}
0 & 0 & 0 & 0  \\
0 &  (1  + 2 \langle j^i_z \rangle )& 2\langle j^i_+\rangle & 0 \\
0 &2 \langle j^i_-\rangle &  (1  - 2 \langle j^i_z \rangle ) & 0 \\
0 & 0 & 0 & 0  
\end{array}
\right] . \label{eq:red1}
\end{eqnarray}    
The von-Neumann  entropy can then be computed  using 
\begin{equation} \label{eq:1nuent}
    S^{(i)}_{1} = - {\rm Tr}\left(R^{(i)}_1 \ln R^{(i)}_1 \right).
\end{equation}  

The reduced one-body density can be expressed also in terms of the polarization components leading to 
\begin{eqnarray}
R^{(i)}_1 & = &
 \left[ 
\begin{array}{cccc}
0 & 0 & 0 & 0  \\
0 & \frac{1}{2} (1 + P_z) & \frac{1}{2} ( P_x + i P_y) & 0 \\
0 & \frac{1}{2} (  P_x - i P_y )& \frac{1}{2} (1 - P_z) & 0 \\
0 & 0 & 0 & 0  
\end{array}
\right]. \label{eq:red1bis}
\end{eqnarray}    
Here, we used the fact that all neutrinos within a beam are equivalent.
The eigenvalues of the density are given by: 
 \begin{eqnarray}
(1- 2 \lambda)^2 = |P|^2 & \longrightarrow \lambda_{\pm} = \frac{1}{2}\left( 1 \pm |P|\right),
\end{eqnarray}
where $|P|^2 = P^2_x + P^2_y+P^2_z$, leading to the expression of the one-neutrino entropy given in Refs.~\cite{Cer19,Pat21}. Equations \eqref{eq:red1} or \eqref{eq:red1bis} each show that the reduced one-body density is directly linked to the expectation
values of one-body observable.   

One can proceed in a similar way to obtain the reduced two-body density $R^{(ij)}_2$ 
associated with two neutrinos in the subsystem $A$ that are labeled by $i$ and $j$. The reduced space associated with the two-neutrino system has a size $2^4$, but, due to symmetries, only 
a $4 \times 4$ block has non-zero components. For the sake of compactness, we only give below the non-zero sub-block of $R^{(ij)}_2$. Following the same 
technique as for the one-body reduced component, a lengthy but straightforward calculation leads to the expression of $R^{(ij)}_2$ in terms of 
the quasi-spin components given by:  
\begin{widetext}
\begin{eqnarray}
R^{(ij)}_2 &=& \frac{1}{4}
\left[ 
\begin{array}{llll}
\langle (1_j + 2 j^j_z)(1_i + 2 j^i_z) \rangle &2\langle (1_j + 2 j^j_z)  j^i_+) \rangle  & 2\langle j^j_+ (1_i + 2 j^i_z)  \rangle & 4 \langle j^j_+ j^i_+  \rangle  \\
2\langle (1_j + 2 j^j_z)  j^i_-) \rangle       &  \langle (1_j + 2 j^j_z)(1_i - 2 j^i_z)  \rangle &  4 \langle j^j_+ j^i_-  \rangle  &  2\langle j^j_+ (1_i - 2 j^i_z)  \rangle  \\
2\langle j^j_- (1_i + 2 j^i_z)  \rangle      & 4 \langle j^j_- j^i_+  \rangle  &\langle (1_j - 2 j^j_z)(1_i + 2 j^i_z)   \rangle&  2\langle (1_j -2  j^j_z) j^i_+   \rangle \\
4 \langle j^j_- j^i_-  \rangle       &  2\langle j^j_- (1_i - 2 j^i_z)  \rangle & 2\langle (1_j - 2 j^j_z) j^i_-   \rangle & \langle (1_j - 2 j^j_z)(1_i - 2 j^i_z) \rangle
\end{array}
\right] . \label{eq:red2}
\end{eqnarray} 
\end{widetext}
Due to the symmetry with respect to the exchange of neutrinos within the subsystem $A$, the above densities are independent of the choices of $i$ or 
$(i,j )$. 
Furthermore, the expectation values entering in the two densities can be related to the mean-values and fluctuations of the $\vec J_A$ components. We have, for instance, for the $z$-component:
\begin{eqnarray} \label{eq:jzJAz}
\langle  j^i_z \rangle = \frac{\langle J_{A,z}\rangle }{N_A}, ~~ \langle  j^i_z j^j_z \rangle = \frac{\langle J^2_{A,z}  \rangle - 1}{N_A(N_A-1)} . \label{eq:1toN}
\end{eqnarray} 

\begin{figure}[htbp] 
\includegraphics[scale=0.3]{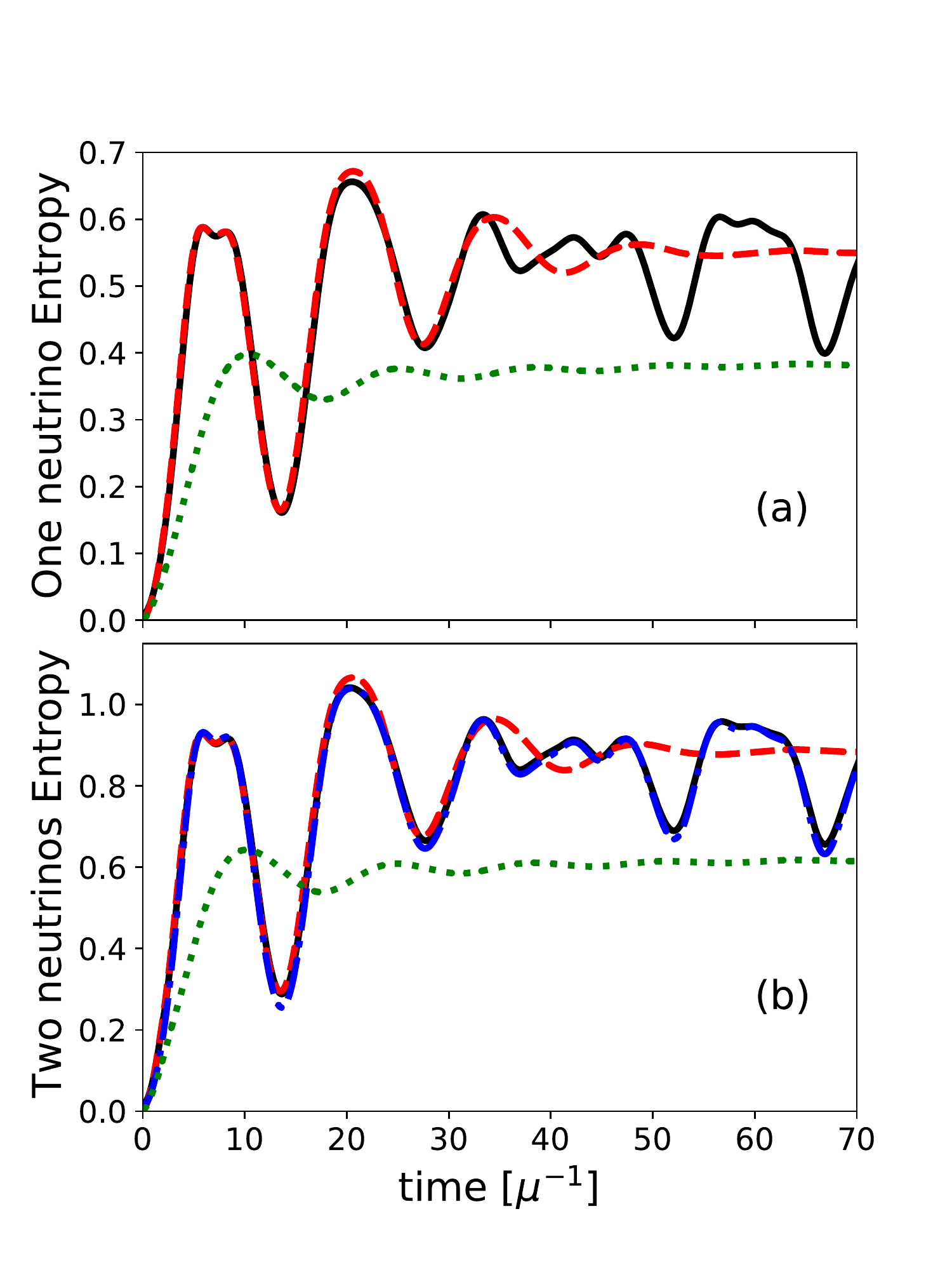} 
    \caption{One-neutrino (a) and two-neutrino (b) entropies obtained as a function of time for the exact (black solid) and approximate phase-space (red dashed) approaches as a function of time. In lower panel, the blue dot-dashed curve represents the exact one-neutrino entropy times a scaling factor of $(\ln 3)/(\ln 2)$. The green dotted curve corresponds to the time-dependent Hamiltonian case discussed in the conclusion.  }
    \label{fig:entrop}
\end{figure}

 The two-neutrino von Neumann entropy is then given by:
\begin{equation} \label{eq:2nuent}
 S^{(ij)}_{2} = - {\rm Tr}\left(R^{(ij)}_2 \ln R^{(ij)}_2 \right). 
\end{equation}
The evolution of the one- and two-neutrino entropy is shown in Fig.~\ref{fig:entrop}. In the mean-field limit, both entropies are zero (and are therefore not shown in the figure). 

The absence of entanglement in mean-field theory is a clear shortcoming of this simplified approach. Equations \eqref{eq:1nuent} and \eqref{eq:2nuent} show how the entanglement between neutrinos is encoded in the components of the reduced one- and two-body density matrices. Furthermore, since the elements of these matrices are related to the expectation values and fluctuations of the one-body observables, one expects that the effects of entanglement will manifest in their evolution, as mentioned before. In particular, we see from the expressions above that a condition for the proper description of the two-body entanglement is the proper account of quantum fluctuations, since the reduced densities are directly expressed in terms of the second moments of the quasi-spin. Equations \eqref{eq:red2}--\eqref{eq:2nuent} also make explicit the link between quantum entanglement and quantum fluctuations (i.e., items (i) and (ii) discussed near the end of the previous section). A corollary to this fact is that the proper description of entanglement could only be achieved by a theory able to describe accurately quantum fluctuations beyond the mean-field picture. 
Such a theory is proposed and discussed in Sec.~\ref{sec:smf}.

 As an aside, we mention one interesting aspect that could be uncovered from Fig. \ref{fig:entrop}. 
We show in panel (b) of this figure that we have the approximate scaling $\xi_{2/1} =S_2^{(ij)}/S_1^{(i)} = (\ln 3)/(\ln 2)$, if both the neutrinos $(i,j)$ are taken from the same beam. This scaling can be explained with the following symmetry argument.

First, assuming that all eigenvalues of $R^{(i)}_1$ and $R^{(ij)}_2$ are equal and completely degenerate would lead to $\xi_{2/1} =2$. This value would represent the most general case where all the sub-components with total quasi-spin 1 (symmetric) and 0 (anti-symmetric) of a composite two-neutrino state are represented. However, in this case, since all the neutrinos within a given beam are assumed to have identical momenta and flavor evolution, only the symmetric subspace is represented. Since this subspace has dimension 3, the ratio of the maximum possible one- and two-neutrino entropies is quenched due to the symmetry constraint. We observe that, even if the entropies are less than maximal, this ratio still represents a reasonable approximation. We checked more generally that the ratio of the $n$-neutrino entropy to the one-neutrino entropy $\xi_{n/1}$, for $n \leq N_A$, is approximately given by $[\ln(n+1)]/[\ln 2]$ and is strongly quenched compared to the symmetry-unrestricted case $\xi_{n/1}=n$.

\section{Husimi phase-space distribution}
\label{sec:husimi}

\begin{figure*}[htbp] 
\includegraphics[scale=0.5]{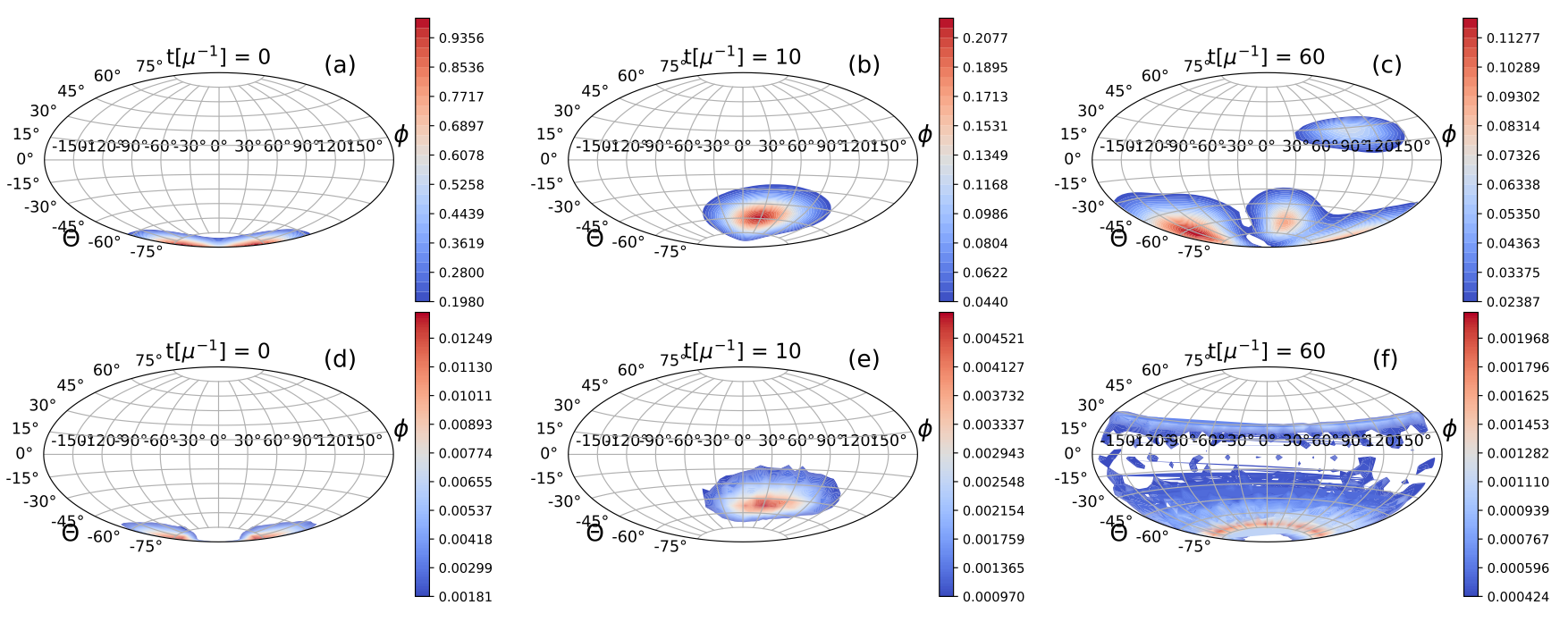}
    \caption{Illustration of the Husimi phase-space distribution obtained at different times $t \mu=0$ (a), $10$ (b), and $60$ (c), for the subsystem $A$.In the top part, we depict the exact solutions.
    In the bottom part, we show the probability distribution directly obtained with the phase-space approach using $10^5$ events (panels (d), (e) and (f)). Note that, in the latter case the probability integrated in bins of $(\theta, \phi)$ is directly normalized to $1$. This normalization factor is different from the Husimi distribution. The results are obtained for the ``bipolar symmetric case" of Ref.~\cite{Mar21} with $\Omega = 0.5$, $\theta_A = \pi - \theta_B$, $\theta_B = 0.2$, $\phi_A = \pi$, $\phi_B=0$ and $N_A=N_B = 50$. Each subfigure shows the phase-space using the Aitoff projection technique that projects a 3D spherical distribution on a 2D space \cite{aitoff}.}
    \label{fig:BipSymHus}
\end{figure*}

We make here a phase-space analysis of the exact subsystem $A$ evolution. We use the Husimi quasi-probability distribution, also called
Q-representation, which has the advantage over the Wigner distribution of being always positive \cite{Hus40,Gar00,Sch01,Lee95}.

\subsection{Husimi distribution for neutrinos}  

To study the phase-space properties, we introduce the Husimi quasi-probability distribution associated with the subsystem $A$. Such a distribution
is obtained by decomposing the reduced density matrix $D_A(t)$ on the over-complete basis formed by coherent states. For the problem considered here, 
these coherent states identify with the coherent states of the $SU(2)$ group \cite{Zha90}. 

Since a phase-space analysis has not been presented previously in the context of collective neutrino oscillations, we recall some important aspects of the Husimi approach that will be useful for the discussion 
below.    
We denote here generically the coherent states for the subsystem $A$ by $| \Omega \rangle$. Such coherent states identify with the Slater determinants  given by Eq.~(\ref{eq:coherent}), with varying 
angles. More precisely, the full set of coherent states are obtained using $| \Omega \rangle = | \theta, \phi \rangle$ in Eq. (\ref{eq:coherent}) with 
$0\leq\theta\leq\pi$ and $0\leq\phi\leq 2\pi$ \cite{Gazeau,Vieira}. We note in passing that the initial state considered previously is one of the coherent 
states with $\theta = \theta_A$ and $\phi =  \phi_A$.   
These coherent states are not orthogonal with each other; in fact 
\begin{eqnarray}
\langle \Omega | \Omega' \rangle &=& \left(\cos \frac{\Theta}{2}\right)^{N_A} \label{eq:ortho}
\end{eqnarray}
with $\cos\Theta = \cos \theta \cos \theta' + \sin \theta \sin \theta' \cos(\phi - \phi')$ \cite{Zha90}. 
These states form an over-complete basis having the closure relation: 
\begin{eqnarray}
 \frac{2J_A+1}{4\pi}\int |\Omega\rangle \langle\Omega| \mathrm{d}\Omega=1, \label{eq:closure}
\end{eqnarray}
with $\mathrm{d}\Omega=\sin\theta\,\mathrm{d}\theta\,\mathrm{d}\phi$.

The Husimi phase-space distribution associated with the density $D_A(t)$ is then defined as:
\begin{eqnarray}
Q_A(\Omega, t) &=& \langle \Omega  | D_A(t) |\Omega\rangle.
\end{eqnarray}
Two useful examples of Husimi quasi-probability distributions are: (i) the one associated with a coherent state itself $|\theta_0,\phi_0\rangle$, 
\begin{align}
 Q(\theta,\phi)\!&=& \!\!\!\left[\frac{1+\cos\theta\cos\theta_0+\sin\theta\sin\theta_0\cos(\phi-\phi_0)}{2}\right]^{N_A}, 
\label{eq:coh}
\end{align}  
and (ii) the one associated with a given $|m_A\rangle$ state: 
\begin{eqnarray}
 Q(\theta,\phi)&=& C^{N_A/2 - m_A}_{N_A} \left(\frac{1+\cos\theta}{2}\right)^{N_A/2+m_A} \nonumber \\
&&\times  \left(\frac{1-\cos\theta}{2}\right)^{N_A/2 - m_A}. 
\label{eq:huseig}
\end{eqnarray} 

The Husimi quasi-probability distribution has the advantage over other phase-space distributions, such as the Wigner function, of being positive for all values of $(\theta, \phi)$. We can also describe the phase-space with a pair of conjugate variables  $p=\cos\theta$ and $q=\phi$, corresponding to 
the normalized relative population difference between the states $1$ and $2$ and the relative phase between these two states, respectively. 
Still, this distribution contains all quantum effects beyond purely the classical limit. Quantum effects are contained in the nonorthogonality of the 
coherent states (see Eq.~(\ref{eq:ortho})). Another signature of the quantum nature of the distribution is that the expectation values 
of any operator $O$ require the introduction of the Weyl symbol denoted hereafter by $P_O(\Omega)$ and defined via:
\begin{eqnarray}
 O =\frac{N_A+1}{4\pi} \int P_O(\Omega) |\Omega\rangle\langle\Omega |\mathrm{d}\Omega.
\label{eq:weyl}
\end{eqnarray}
In particular, we have for the expectation value of any operator $O$: 
\begin{eqnarray}
{\rm Tr}_A \left[ O D_A(t)\right] &=& \frac{N_A+1}{4\pi} \int P_O(\Omega) Q_A(\Omega) \mathrm{d}\Omega.  \label{eq:husimimean}
\end{eqnarray}    
For instance, the Weyl symbols for the spin operators $J^A_{x,y,z}$ are given by \cite{Yil14}:
\begin{eqnarray}
 P_{J^A_\alpha }(\Omega)&=&\frac{N_A+2}{N_A}\langle\Omega |J^A_\alpha|\Omega\rangle,\label{eq:weyl-spin}
 \end{eqnarray}
 with
\begin{eqnarray}
\left\{
\begin{array}{l}
\displaystyle \langle J^A_x \rangle_\Omega = \frac{N_A}{2}\sin\theta\,\cos\phi,  \\
\\
\displaystyle \langle J^A_y \rangle_\Omega=\frac{N_A}{2}\sin\theta\,\sin\phi, \\
\\
\displaystyle \langle J^A_z \rangle_\Omega=\frac{N_A}{2}\cos\theta, 
 \end{array}
\right. 
\label{eq:jtheta}
\end{eqnarray}
where we used the compact notations $\langle J^A_\alpha \rangle_{\Omega} = \langle\Omega| J^A_\alpha | \Omega\rangle$. We can similarly obtain for the non-centered moments of the quasi-spins: 
 \begin{eqnarray}
 P_{\{ J^A_\alpha , J^A_\beta  \}/2}(\Omega)&=&\frac{(N_A+2)(N_A+3)}{N^2}\langle J^A_\alpha \rangle_{\Omega} \langle J^A_\beta \rangle_{\Omega} \nonumber \\
&-& \frac{N+2}{4}\delta_{\alpha \beta}, 
\label{eq:pipj}
\end{eqnarray}  
where $\{  J^A_\alpha , J^A_\beta \} = J^A_\alpha J^A_\beta + J^A_\beta  J^A_\alpha$. 

Let us consider the polarization components of the system $A$ given by $P_{A,\alpha} = 2  \langle J^A_\alpha \rangle/N_A $. 
Denoting $2 \langle J^A_\alpha \rangle_{\Omega} / N_A \equiv P_{A,\alpha}(\Omega) $, then due to Eq. (\ref{eq:jtheta}), we have for each coherent state:
\begin{eqnarray}
P^2_{A,x}(\Omega) + P^2_{A,y} (\Omega)  + P^2_{A,z}(\Omega) = 1 .
\end{eqnarray}   
One can therefore assign to each coherent state a point on the Bloch sphere corresponding to the crossing between the line defined by the vector 
$\vec P_\alpha (\Omega)$ and the Bloch sphere of radius $1$. However, it should be kept in mind that these coherent states are also described by a certain dispersion in phase-space, given by Eq.~(\ref{eq:coh}). 
This dispersion together with the 
use of non-trivial Weyl symbols prevent a direct interpretation of the Husimi distribution as a classical probability. Nevertheless, in the limit of large 
particle numbers $N_A \rightarrow + \infty$, we see from Eq.~(\ref{eq:ortho}) that we have $\langle \Omega | \Omega' \rangle \rightarrow \delta(\Omega - \Omega')$. We also observe from Eqs.~(\ref{eq:weyl-spin}) and (\ref{eq:pipj}) the limits:  
\begin{eqnarray}
 P_{J^A_\alpha }(\Omega) &\longrightarrow&  \langle\Omega |J^A_\alpha|\Omega\rangle, \nonumber \\
 P_{\{ J^A_\alpha , J^A_\beta  \}/2}(\Omega) &\longrightarrow& P_{J^A_\alpha }(\Omega)  P_{J^A_\beta}(\Omega) . \nonumber
 \end{eqnarray}
More generally, symmetrized moments of any combination $\{ J^A_{\alpha_1} , \cdots , J^A_{\alpha_k} \}$ will identify with the 
product $ P_{J^A_{\alpha_1}} (\Omega) \cdots  P_{J^A_{\alpha_k}} (\Omega)$.  Therefore, in the large $N_A$ limit, Eq.~(\ref{eq:husimimean}) 
identifies with the standard statistical average in classical theory and the Husimi distribution can be interpreted with no ambiguity as a classical probability.

\subsection{Exact phase-space evolution for the two neutrino beam problem}


Starting from the reduced density $D_A(t)$ given by Eq. (\ref{eq:dat}) and obtained by solving exactly the two neutrino beams problem, we
computed explicitly the Husimi distribution as a function of time. In practice, this calculation is achieved starting from  Eq.~\eqref{eq:dat} and using Eq.~\eqref{eq:huseig} 
for each state $| m_A \rangle$. Illustrations of the subsystem $A$ Husimi distribution are shown at different times in Fig.~\ref{fig:BipSymHus} for the Bipolar symmetric case.  

We observe in Fig.~\ref{fig:BipSymHus}a-b that the Husimi distribution is rather localized for short time evolution. This trend is indeed expected, since we assume that the initial conditions for subsystem $A$ (or $B$) are both coherent states. For the system $A$ (resp.~$B$) the initial Husimi distribution
therefore identifies with Eq.~\eqref{eq:coh} provided that $(\theta_0, \phi_0)$ are replaced by $(\theta_A, \phi_A)$ (resp. $(\theta_B, \phi_B)$ and $N_A$ is replaced by $N_B$). For a large enough neutrino number as considered in the present example, Eq.~\eqref{eq:coh} verifies:
\begin{eqnarray}
\langle \Omega | \Omega' \rangle &\simeq&  e^{-N_A \Theta^2/8}
\end{eqnarray} 
and the distribution associated with a coherent state identifies with a localized Gaussian distribution in phase-space. Such a localized distribution is 
the one shown in Fig.~\ref{fig:BipSymHus}a. In the mean-field approximation, the wave-function is assumed to remain coherent during the evolution. 
This assumption automatically implies a Gaussian approximation for the phase-space distribution together with the impossibility of describing 
large, complicated fluctuation patterns. 
For long time evolution (panel (c)),  the phase-space distribution has a multimodal structure with several localized peaks, unambiguously revealing the non-Gaussian nature of the reduced evolution. These highly nontrivial large fluctuations emerge due to the coupling and entanglement with the subsystem $B$, and are much beyond the effects that could be treated in a mean-field framework.   

In view of Fig.~\ref{fig:BipSymHus}c, one can anticipate that even when beyond mean-field effects are included, 
it is unlikely that a method based on a Gaussian approximation \cite{Wan66,Ler20,Rog22} can account for this complex behavior. 
For instance, a standard strategy to go beyond the mean-field approximation is to use the BBGKY hierarchy and truncate the equations of motion at second or higher order. At second order, this truncation is equivalent to following the first and second moments of the $\vec J_{A/B}$ components \cite{Bon16,Vol13,Vol15}. However, we conclude from the complexity of the distribution shown in Fig.~\ref{fig:BipSymHus} that the accurate description of such distribution can only be achieved if higher moments are included as well as their actions on first and second moments.

\section{Approximate phase-space method for neutrino oscillations}
\label{sec:smf}

In this section, we propose a method to accurately describe the exact evolution presented previously. To obtain an approximate description of the entanglement evolution of two neutrino beams, 
we have adapted here the phase-space approach (PSA) originally proposed in Refs.~\cite{Ayi08,Lac14} to the  neutrino oscillations problem. 
This method was successfully applied to different areas of physics \cite{Lac12,Lac13,Yil14,Lac14b,Ulg19,Czu20} and was shown to be rather accurate while not relying on any Gaussian approximation of the phase-space distribution \cite{Lac16}.   
In PSA, the quantum problem of interacting fermions is mapped into a statistical problem where an ensemble of initial conditions is considered. The initial fluctuating conditions are chosen in such a way that 
the classical average over the initial values matches the quantum expectation of the initial state. Then, each initial condition is evolved using mean-field  
EOMs that are independent of each other. The fact that only the mean-field evolution is needed makes the method rather simple and versatile. We discuss below how the approach can be adapted to the neutrinos case.

\subsection{Matching initial conditions for the two-beam problem}

The PSA replaces the exact many-body problem by an ensemble
of independent mean-field trajectories with fluctuating initial conditions \cite{Ayi08, Lac14}. For the present model, this corresponds to considering a distribution of initial values for the polarization components that will then be used to solve the time-dependent equation (\ref{eq:mf}). These initial conditions are treated as classical configurations of our system, restricted by the crucial property that their statistical averaging 
exactly reproduces the mean values and quantum fluctuations of the polarization 
obtained with the initial wave function we are given. 
In this sense, the PSA approach replaces a quantum problem by a statistical problem and the observables' evolution is obtained by performing classical statistical averages 
over different trajectories. We detail below how fluctuating initial conditions reproducing quantum expectations values are constructed.  

The initial many-body state considered in this work corresponds to a tensor product  $| \Psi \rangle = | \Omega_A \rangle \otimes | \Omega_B \rangle$, where both states are coherent states (see Eqs.~(\ref{eq:coherent}-\ref{eq:coherent2})). Because of the tensor product form of the initial state, one can consider the fluctuations in subsystems $A$ and $B$ separately. 

 We focus here first on the subsystem $A$, with the discussion being identical for subsystem $B$. 
 The mean values and fluctuations are easier to compute in the rotated basis where, for each neutrino, the operator $a^\dagger_{A,i}$ given by Eq.~\eqref{eq:coherent2} is complemented by the creation operator:  
\begin{eqnarray}
a^\dagger_{\bar A,i} &=& -  \sin \left( \frac{\theta_{A}}{2}\right) e^{ - i \phi_{A}}  a^\dagger_{1,i} + \cos \left( \frac{\theta_{A}}{2}\right)a^\dagger_{2,i} .\nonumber 
\end{eqnarray}
We note in passing that the creation operators $(a^\dagger_{A,i} , a^\dagger_{\bar A,i})$ correspond to the flavor basis. We introduce 
the associated quasi-spin operators $({\cal J}^A_x, {\cal J}^A_y, {\cal J}^A_z )$.  The state $| \Omega_A \rangle$ corresponds
to the lowest eigenstates of ${\cal J}^A_z$ in the rotated space with eigenvalue $m_A = -N_A/2$. This observation immediately gives us:
\begin{eqnarray}
\langle {\cal J}^A_x \rangle &=&  \langle {\cal J}^A_y \rangle  = 0, ~~\langle {\cal J}^A_z \rangle = -\frac{N_A}{2}. \nonumber 
\end{eqnarray}
In the following, we define the quantum second moment of two operators acting in the $A$ space, denoted by
$X$ and $Y$, as:
\begin{eqnarray}
\Sigma^2_{A, XY} &=& \frac{1}{2} \langle X Y + Y X \rangle - \langle X\rangle   \langle Y\rangle,  
\end{eqnarray}
where the expectation values are performed over the total system. 
It is straightforward to show that the second moments of the ${\vec {\cal J}}$ components at the initial time are given by
\begin{eqnarray}
\Sigma^2_{A, xz} &=& \Sigma^2_{A,yz} = \Sigma^2_{A,xy} = \Sigma^2_{A,zz} =0.
\end{eqnarray}
The only non-zero fluctuations are $\Sigma^2_{A,xx}$ and $\Sigma^2_{A,yy}$, for which we have \cite{Lac12}:
\begin{eqnarray}
 \Sigma^2_{A, xx} &=& \Sigma^2_{A, yy} = \frac{N_A}{4}. 
\end{eqnarray}

We consider now the PSA method. In this approach, we introduce a statistical 
ensemble of values $\left({\cal J}^{A(\lambda)}_x, {\cal J}^{A(\lambda)}_y, {\cal J}^{A(\lambda)}_z \right)$ where $\lambda=1, \ldots, N_{\rm evt}$ labels 
the events. These variables are treated as classical objects. Then, mean values and fluctuations are obtained by classical averages over the events. A simple 
way to reproduce the quantum means and second moments given above is to assume that ${\cal J}^{A(\lambda)}_z  = -N_A/2$ is a non-fluctuating 
variable while ${\cal J}^{A(\lambda)}_x$ and ${\cal J}^{A(\lambda)}_y$ are considered as Gaussian Stochastic variables with mean zero and 
widths equal to $N_A/4$.   

The mean-field equations that will be used for the evolution are given in the mass basis. Therefore, it is necessary to transform the initial fluctuations in the rotated space to the fluctuations in the original space where the EOMs~\eqref{eq:mf}
are written. To do so, we use the fact that the two sets of quasi-spin quantum operators are linked through:
\begin{widetext}
\begin{eqnarray}
J^A_{x} &=& +\left[ c^2_A -s^2_A \cos(2 \phi_A) \right] {\cal J}^A_x - s^2_A \sin(2 \phi_A) {\cal J}^A_y+ \sin(\theta_A) \cos(\phi_A) {\cal J}^A_z , \label{eq:jxinv} \\
J^A_{y} &=& -s^2_A   \sin(2 \phi_A) {\cal J}^A_x + \left[ c^2_A +s^2_A \cos(2 \phi_A) \right] {\cal J}^A_y + \sin(\theta_A) \sin(\phi_A) {\cal J}^A_z  , \label{eq:jyinv} \\
J^A_{z} &=& -  \sin(\theta_A) \cos(\phi_A) {\cal J}^A_x - \sin(\theta_A) \sin(\phi_A) {\cal J}^A_y + \cos(\theta_A) {\cal J}^A_z  ,\label{eq:jzinv}
\end{eqnarray}
with $c_A = \cos(\theta_A/2)$ and $s_A = \sin(\theta_A/2)$. 
It is easy to check that, if we replace the operators by the fluctuating quasi-spin $\left({\cal J}^{A(\lambda)}_x, {\cal J}^{A(\lambda)}_y, {\cal J}^{A(\lambda)}_z \right)$ in the right-hand side, we obtain a new set of fluctuating quantities $\left({J}^{A(\lambda)}_x, {J}^{A(\lambda)}_y, {J}^{A(\lambda)}_z \right)$ that will reproduce properly the quantum fluctuations in non-rotated space.      

If, instead of quasi-spin, we use the polarization vectors and finally obtain:
 \begin{eqnarray}
P^{A(\lambda)} _x &=& +\left[ c^2_A -s^2_A \cos(2 \phi_A) \right] {\cal P}^{A(\lambda)}_{x} - s^2_A \sin(2 \phi_A) {\cal P}^{A(\lambda)}_{y} + \sin(\theta_A) \cos(\phi_A) {\cal P}^{A(\lambda)}_{z},  \nonumber \\
P^{A(\lambda)} _y  &=& -s^2_A   \sin(2 \phi_A) {\cal P}^{A(\lambda)}_{x} + \left[ c^2_A +s^2_A \cos(2 \phi_A) \right] {\cal P}^{A (\lambda)}_{y} + \sin(\theta_A) \sin(\phi_A){\cal P}^{A(\lambda)}_{z}, \nonumber  \\
P^{A(\lambda)} _z  &=& -  \sin(\theta_A) \cos(\phi_A) {\cal P}^{A(\lambda)}_{x}- \sin(\theta_A) \sin(\phi_A) {\cal P}^{A(\lambda)}_{y} + \cos(\theta_A) {\cal P}^{A(\lambda)}_{z},\nonumber 
\end{eqnarray}
where ${\cal P}^{A(\lambda)}_{z} = 1$ is constant for all events, while ${\cal P}^{A(\lambda)}_{x} $ and ${\cal P}^{A(\lambda)}_{y} $ are two random Gaussian numbers with mean zero and variances equal to $1/N_A$. 

The initial fluctuations of subsystem $B$ can be obtained in a similar way, leading to two additional Gaussian random numbers ${\cal P}^{B(\lambda)}_{x} $ and ${\cal P}^{B(\lambda)}_{y} $ with variances equal to $1/N_B$. 
\end{widetext}

\subsection{Simulation of the evolution by independent mean-field paths}

In the PSA approach, a set of initial fluctuating conditions is used for the polarization components $(\vec P^{(\lambda)}_A, \vec P^{(\lambda)}_B)$ where $\lambda = 1,\cdots, N_{\rm evt}$. 
Each initial condition (event) $\lambda$ is evolved independently from the others according to the mean-field equation, i.e.:  
\begin{eqnarray}
\left\{ 
\begin{array}{l}
\displaystyle  \dot P^{A(\lambda)}_x =  + \frac{\Omega}{2}  P^{A(\lambda)}_y + x_B C^{AB(\lambda)}_x   \\  
\\
\displaystyle  \dot P^{A(\lambda)}_y =  - \frac{\Omega}{2}  P^{A(\lambda)}_x +  x_B C^{AB(\lambda)}_y   \\  
\\
\displaystyle  \dot P^{A(\lambda)}_z =  +  x_B C^{AB(\lambda)}_z    \\
\\
\displaystyle  \dot P^{B(\lambda)} _x =  - \frac{\Omega}{2}  P^{B(\lambda)} _y - x_A C^{AB(\lambda)}_x   \\  
\\
\displaystyle  \dot P^{B(\lambda)} _y =  + \frac{\Omega}{2}  P^{B(\lambda)} _x -  x_A C^{AB(\lambda)}_y   \\  
\\
\displaystyle  \dot P^{B(\lambda)} _z =  -  x_A C^{AB(\lambda)}_z    
\end{array}
\right. 
\end{eqnarray}

In practice, at a given time, the mean-value and second moments of a given observable are deduced by performing the 
classical average of this quantity. We have for instance for the mean polarization and its fluctuation the definition (for $\alpha=x,y,z$):
\begin{eqnarray}
\overline{P^{A(\lambda)}_{\alpha} [t]} &=& \frac{1}{N_{\rm evt}} \sum_{\lambda} P^{A(\lambda)}_{\alpha} [t], \nonumber  \\
\sigma^2_\alpha [t] &=& \overline{\left\{ P^{A(\lambda)}_{\alpha}  [t] \right\}^2} -  \overline{P^{A(\lambda)}_{\alpha}  [t]} ^2 \nonumber 
\end{eqnarray}     

Results obtained with the PSA approach are shown in Fig.~\ref{fig:BipSymP123} and further illustrated in Appendix \ref{app:example}. 
In all examples considered, the method successfully reproduces the average evolution and quantum fluctuation beyond mean-field, although it misses the long-term oscillations of the fluctuations. 

As shown in Ref.~\cite{Lac16}, one key ingredient of this approach is that it accounts for higher order moments of observables without any specific truncation scheme. An illustration of the phase-space explored by the trajectories is given in Fig.~\ref{fig:BipSymHus}d-f. We observe that the long time evolution (panel (f)) presents differences with the exact evolution. The PSA approach does not properly describe the localization along the $\phi$-axis while the splitting along the $\theta$-axis is reproduced to some extent. Despite these differences, Fig.~\ref{fig:BipSymP123} clearly demonstrates its predictive power for mean values and fluctuations.       

From the mean values and fluctuations of the polarization obtained by averaging statistically over trajectories, one can finally construct the equivalent of 
the one- and two-neutrino densities and evaluate the corresponding entropies. The results are shown in Fig.~\ref{fig:entrop}, where a good agreement with the exact results is observed. We tested extensively the PSA approach by considering the various sets of 
initial angles $(\theta_{A/B}, \phi_{A/B})$ like those reported in Ref.~\cite{Mar21} and always obtained very satisfactory results (see Appendix \ref{app:example}).

\subsection{Application to time-dependent Hamiltonian}
\label{sec:timedep}

We have shown above that the PSA approach is predictive for the two-neutrino beam with a time-independent Hamiltonian. As underlined in the introduction, 
one of the difficulties in describing neutrino oscillations subject to multi-beam entanglement is related to the fact that the Hamiltonian is \emph{time-dependent}. 
This complication introduces obstacles in the application of most numerical techniques able to treat the neutrino interaction. Of the semi-analytic and numerical methods in literature that have been used to treat this problem, some can only be applied in the time-independent or adiabatic limit (e.g,. Bethe Ansatz methods~\cite{Peh11,Bir18,Pat19,Cer19}, or exact methods for two-beam systems~\cite{Xio21,Mar21}); those that can go beyond the adiabatic regime, such as Runge-Kutta~\cite{Rra19,Pat21} or Tensor Network-based~\cite{Rog21,Cer22} numerical integration, are currently able to treat only a limited number of neutrinos [$\mathcal{O}(10\text{--}100)$ neutrinos depending on the symmetries in the system]. The adiabatic picture is expected to hold rather far from the compact object from which neutrinos are emitted but may not be valid close to the point of emission. 

For such types of problems, the PSA approach can be a very useful tool, since it is simple to implement, it is predictive, and non-adiabatic effects are automatically included through the mean-field evolution. To illustrate such a situation, we follow Ref.~\cite{Dua06,Cer19,Pat21,Cer22,Bir18} and consider a time-dependent Hamiltonian where the two-body part of the Hamiltonian given by 
Eq.~(\ref{eq:2beam}) is multiplied by a time-dependent factor: 
\begin{eqnarray}
F(t) = [ 1 - \left( 1 - (R^2_\nu / r(t))^2 \right)^{1/2}]^2.  \label{eq:td}
\end{eqnarray}
Such parameterization mimics the weakening of the two-neutrino interaction with the distance from the source of emission.  Here, $R_\nu$ stands for the emitter radius and is taken as $R_\nu= 32.2 \:\Omega^{-1}$ \cite{Pat21}.
$r(t)$ should be interpreted as distance from the center of the emitter to the point of interest at which neutrinos are interacting with each other. This distance ranges 
\emph{a priori} from $r(0) = R_\nu$ to infinity. One source of difficulty is that $F(t)$ varies rapidly when $r(t)$ is close to $R_\nu$ 
and non-adiabatic effects are expected to be important.  

In Ref.~\cite{Cer19}, a Bethe-Ansatz approach, able to treat many-body entanglement 
in the adiabatic regime, was applied to the neutrino entanglement problem where the distance $r(t)$ was parametrized as $r(t) = r_0 + t$, 
with $r_0$ the initial value of $r(t)$. Because of the adiabatic assumption, the method proposed in Ref. \cite{Cer19}, was applied to cases where 
$r_0 =  210.64 \:\Omega^{-1} \gg R_\nu$, i.e. already rather far from the neutrino emission point. With the PSA approach, we do not have this limitation 
and can consider a more realistic situation where $r_0 = R_\nu= 32.2 \:\Omega^{-1}$.  Results obtained with the time-dependent 
Hamiltonian in the non-adiabatic regime are reported with dotted green lines in Figs.~\ref{fig:BipSymP123} and \ref{fig:entrop}. 
We observe that the use of time-dependent two-body interaction affects the evolution significantly. This difference is actually expected due to the reduction of the two-body interaction induced by Eq.~(\ref{eq:td}) as time increases. The two-body entanglement entropy is also reduced compared to the case where the strength of the interaction is fixed to its initial value. With this example, we illustrate that the PSA approach we propose in the present work will be useful 
to study both qualitatively and quantitatively neutrinos oscillation with time-dependent coupling.         

\section{Conclusions and Discussion}

We studied here the connections between the dispersion in phase-space and the entanglement entropy for two interacting neutrino beams.
The interaction between neutrinos entails large non-Gaussian fluctuations in phase-space leading to a non-trivial evolution of the entanglement between the two subsystems. The Husimi distribution computed here clearly underlines the need to properly describe quantum fluctuations beyond the second moments in phase-space. 

We propose an approach, called the phase-space approach (PSA), where a set of independent mean-field trajectories with random initial conditions accurately describes the gross features of entanglement between neutrino beams. This approach, illustrated here for two beams, turns out to 
provide a good reproduction of both one-body and two-body evolution to describe neutrino oscillations including the effect of coupling between different neutrino beams. 

Due to its simplicity, the PSA approach can for instance easily be generalized to many beams having various neutrino numbers in each beam and evolving through a time-dependent interaction Hamiltonian. An illustration of application to the case of time-dependent Hamiltonian is made 
for a situation where the adiabatic assumption is expected to break down.    
      
\section*{Acknowledgments}

We thank... This project has received financial support from the CNRS through
the 80Prime program and is part of the QC2I project. 
It was supported in part by the U.S.~Department of Energy, Office of Science, Office of High Energy Physics, under Awards No.~DE-SC0019465 and DE-FG02-95ER40907 and in part by the U.S. National Science Foundation Grants No.~PHY-2020275 and PHY-2108339. The work of A.~V.~P. was supported by the U.S. Department of Energy under contract number DE-AC02-76SF00515.

\appendix

\section{Additional illustrations of exact, mean-field and PSA results}
\label{app:example}

\begin{figure}[htbp] 
\includegraphics[scale=0.35]{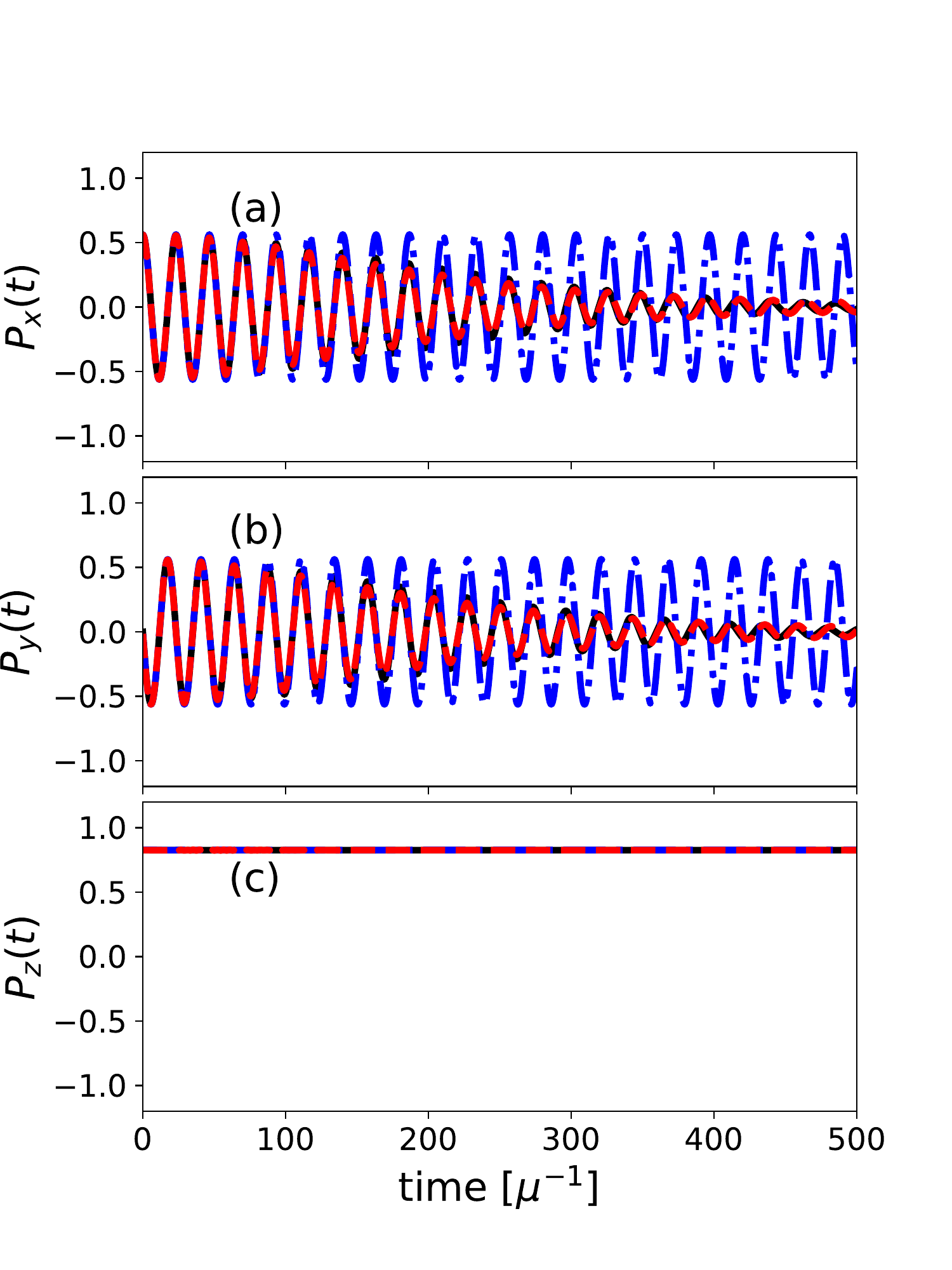}
\includegraphics[scale=0.35]{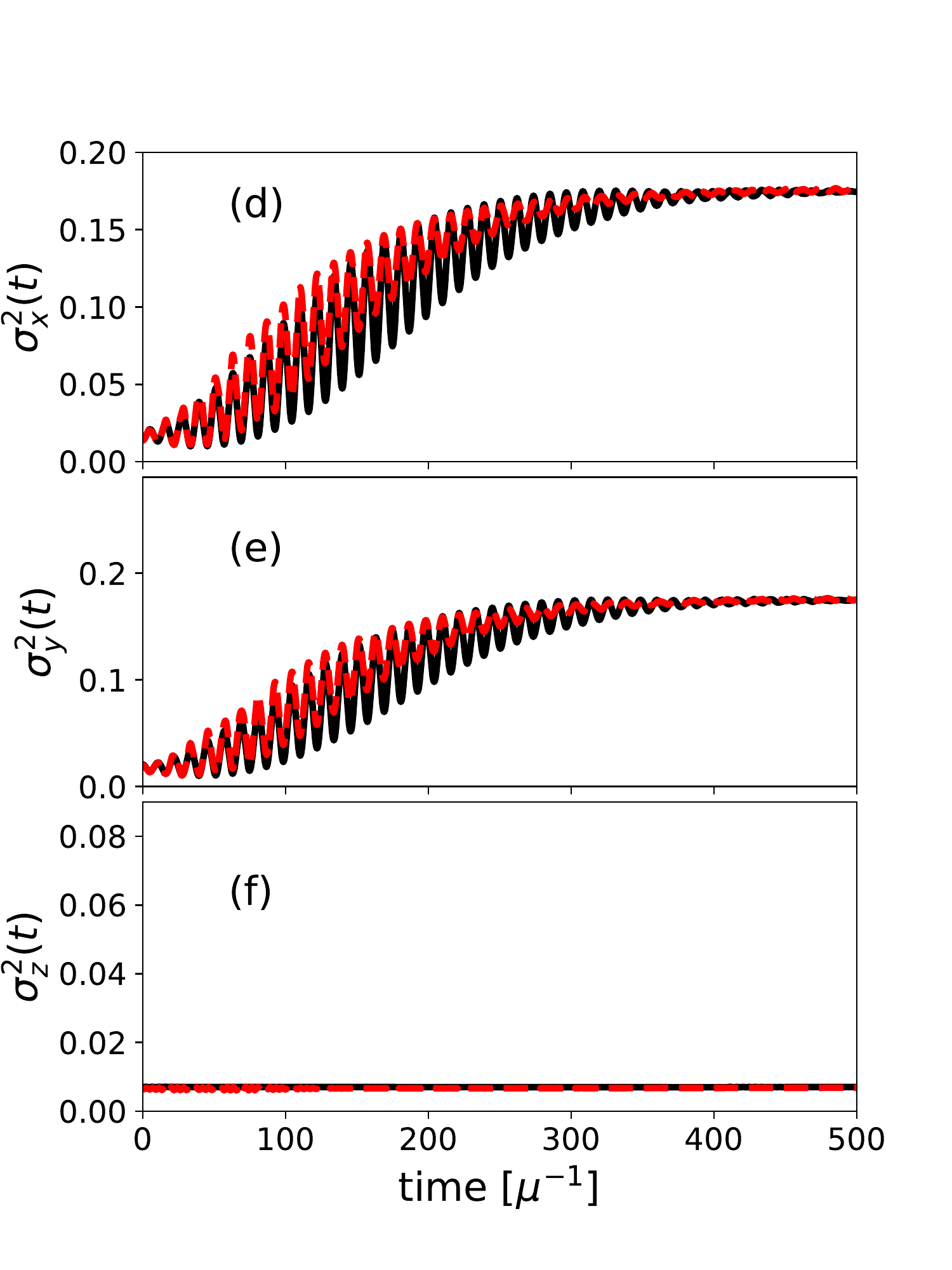} 
    \caption{Illustration of the evolution of the polarization components $\vec P_A = 2 \langle \vec J_A \rangle/N_A$ (left) and their quantum fluctuation (right) for the Precession Symmetric case  with $N_A=N_B =50$ as a function of time for the exact (black solid line), mean-field (blue dot-dashed line) and phase-space method (red dashed line). In the right panel, the mean-field fluctuations are constant in time and remains equal to their initial values (not shown). }
    \label{fig:PrecSymP123}
\end{figure}

\begin{figure}[htbp] 
\includegraphics[scale=0.3]{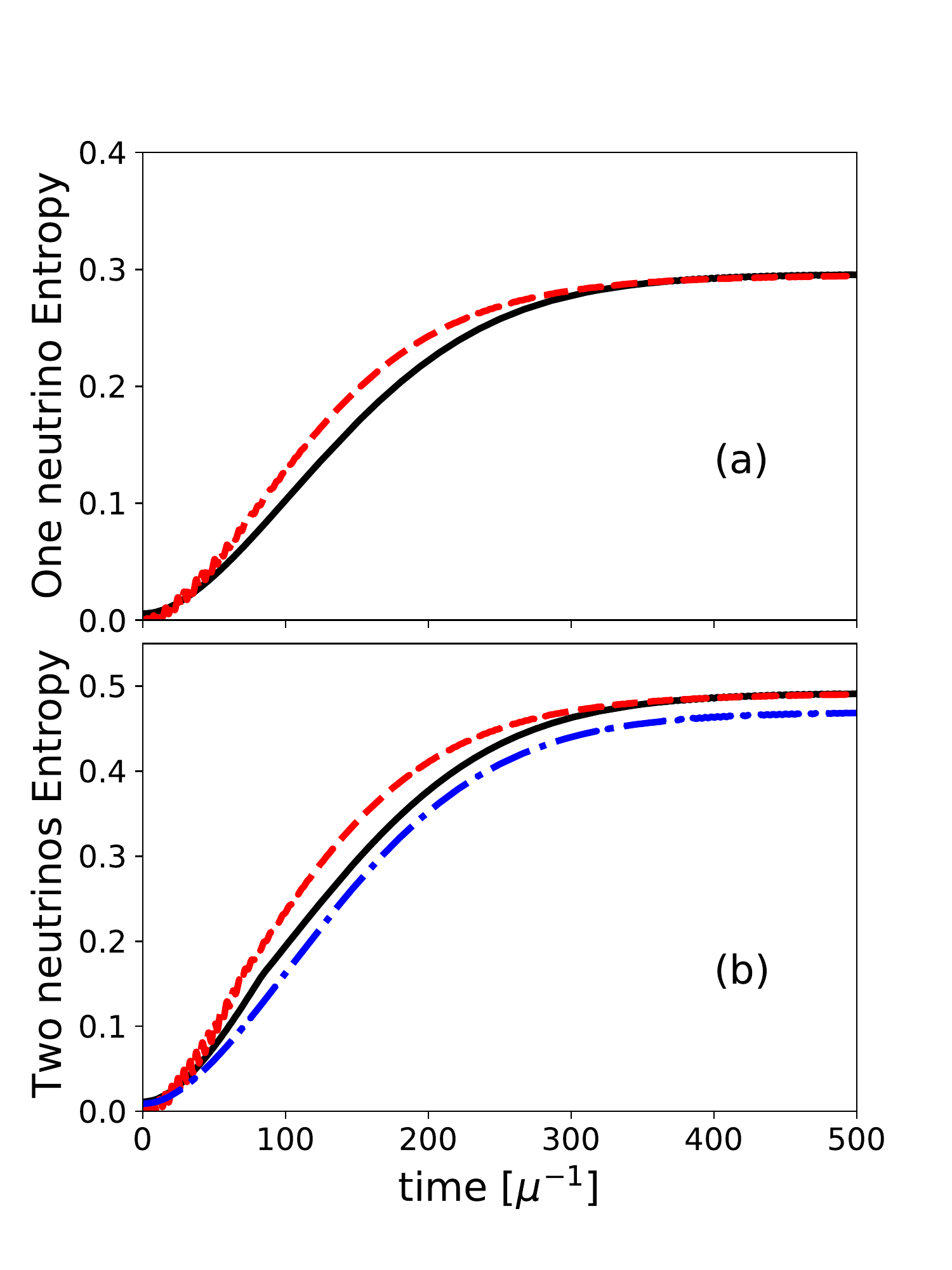} 
    \caption{One neutrino (a) and two-neutrino (b) entropies obtained as a function of time for the exact (black solid line) and approximate phase-space (red dashed line) approaches as a function of time for the symmetric Precession case. 
    In lower panel, the blue dot-dashed line represents the one-neutrino entropy multiplied by the factor $\ln(3)/\ln(2)$. }
    \label{fig:entropprec}
\end{figure}

\begin{figure}[htbp] 
\includegraphics[scale=0.35]{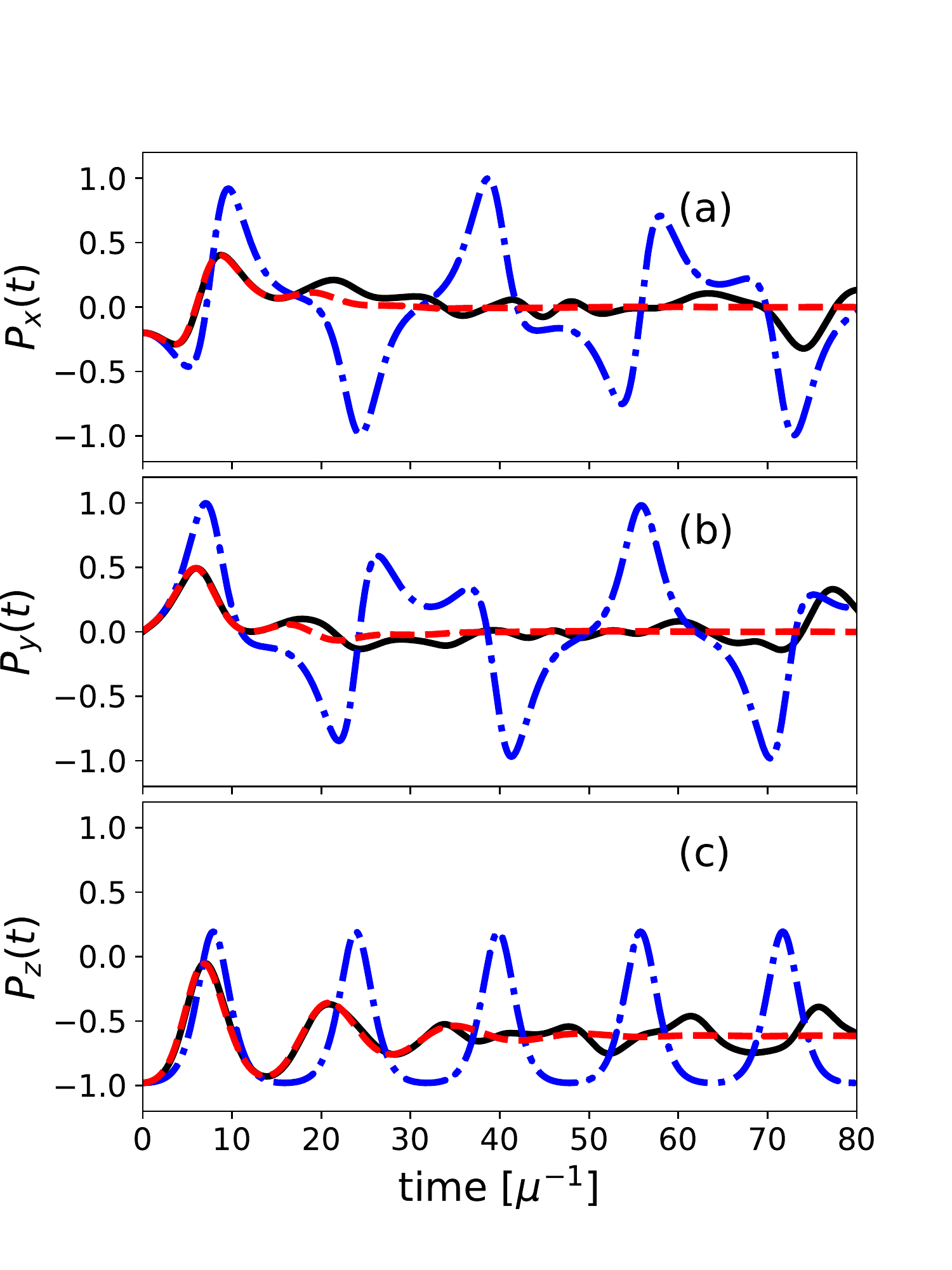}
\includegraphics[scale=0.35]{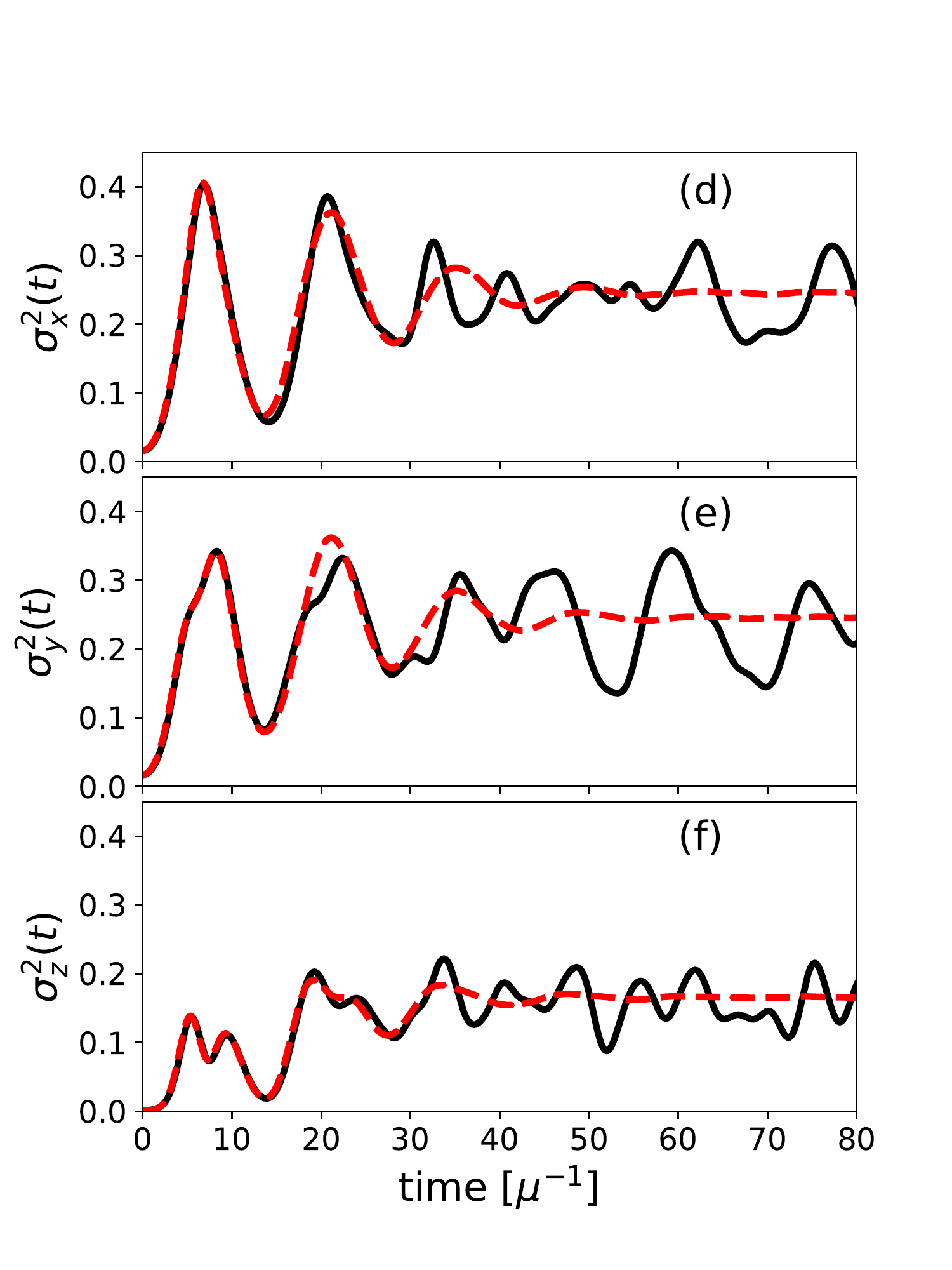} 
    \caption{Same as Fig.~\ref{fig:PrecSymP123} for the asymmetric Bipolar case with $N_A=60$ and $N_B=40$.}
    \label{fig:BipasymP123}
\end{figure}

\begin{figure}[htbp] 
\includegraphics[scale=0.3]{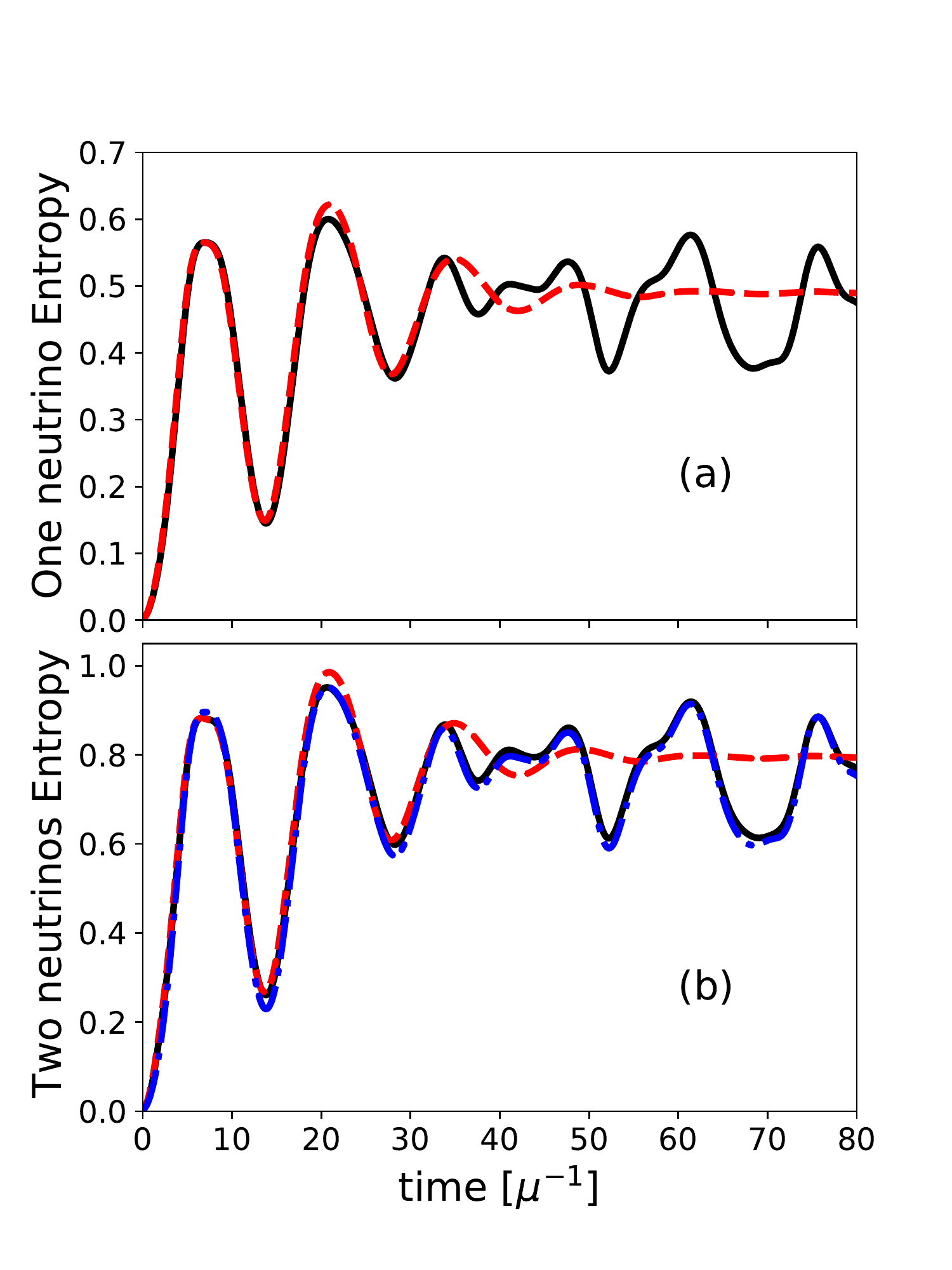} 
    \caption{Same as Fig.~\ref{fig:entropprec} for the asymmetric Bipolar case with $N_A=60$ and $N_B=40$.  }
    \label{fig:entropbipolasym}
\end{figure}

We made extensive applications of both the exact Husimi 
quasi-probability distribution and comparisons with the phase-space approach by picking several examples of initial conditions 
in the tables of Ref.~\cite{Mar21} including symmetric $N_A = N_B$ or asymmetric $N_A \neq N_B$ situations. In all cases, we found very good 
agreement between the exact evolution. We illustrate in Fig.~\ref{fig:PrecSymP123} the evolution of the mean polarization and of its second moments for the ``Precession mode'' case with the initial condition reported in Table 
\ref{tab:paramcte}. The corresponding entropies are shown in Fig.~\ref{fig:entropprec}. Finally we show an asymmetric case ($N_A \neq N_B$) 
for the bipolar parameters in Figs.~\ref{fig:BipasymP123} and \ref{fig:entropbipolasym}. 

\end{document}